\documentstyle[12pt]{article}
\begin{document}
\begin{titlepage}
\hspace{9cm} SISSA/ISAS/100/93/EP

\vspace{2.5cm}
\begin{centering}

{\huge Quantum mechanics and quantum Hall effect
on Riemann surfaces}\\
\vspace{1cm}
{\large Roberto Iengo$^{a,b}$
and Dingping Li$^a$\\
International School for Advanced studies, SISSA, I-34014
Trieste, Italy$^a$\\
Istituto Nazionale di Fisica Nucleare, INFN, Sezione di Trieste,
Trieste, Italy$^b$} \\
\end{centering}
\vspace{.25cm}
\begin{abstract}

The quantum mechanics of a system of charged
particles interacting with a magnetic field
on Riemann surfaces is studied.
We explicitly construct the wave functions of ground states
in the case of a metric proportional to the Chern form
of the $\theta$-bundle,  and the wave functions of
the Landau levels in the case of the the Poincar{\' e}  metric.
The degeneracy of the the Landau levels is obtained by using
the Riemann-Roch theorem. Then we construct the Laughlin wave function
on Riemann surfaces and discuss the mathematical structure
hidden in the Laughlin wave function.
Moreover  the degeneracy of the Laughlin states is also
discussed.

\end{abstract}
\vspace{.25cm}

\end{titlepage}
\def\carre{\vbox{\hrule\hbox{\vrule\kern 3pt
\vbox{\kern 3pt\kern 3pt}\kern 3pt\vrule}\hrule}}

\pagebreak

\section{Introduction}

We study here the quantum mechanics of a system of charged particles living
on a two dimensional surface and interacting with a magnetic field
orthogonal to the surface. When the surface is an infinite plane and the
magnetic field is constant, we have  the well known problem of the
Landau levels.  This problem has received a renewed interest in the context
of the quantum Hall effect (QHE)(for a review, see Ref.\ \cite{pg}).
In fact, the QHE appears to be related to a rich physical
and also mathematical structure,
which is worthwhile to investigate
in various possible configurations.

A particular intriguing and
interesting case occurs when the two dimensional
surface is a Riemann surface of high genus.
Although not directly accessible to experiments, the problem
of the physics on the Riemann surface happens to have deep relations
with modern investigations on some interesting problems, like
the occurrence of chaos in the surface with a negative curvature
\cite{gutzw}, and recent developments
in the theory of Riemann surfaces,
for example, the moduli of
the surface and the vector bundles defined on
the moduli \cite{fay}.

In this paper, we will explore the problem directly from  the
point of view of quantum mechanics,
by defining the Hamiltonian and then constructing
eigenfunctions of the Hamiltonian.
In Ref.\ \cite{comtet}, the problem in the case of
the surface being an open hyperbolic plane
with a constant negative curvature
was considered and Ref.\ \cite{antoine}
discussed the scattering on
a hyperbolic torus.
Instead, here we will mainly consider
compact Riemann surfaces.
Ref.\ \cite{avron} obtained
some interesting results about eigenvalues
and their multiplicity of a particle interacting with the
magnetic field,  in the case of Riemann  surfaces of high genus with
a constant curvature by using some results from the mathematical
literature, for example by using Selberg trace formula
(for a review on Selberg trace formula, see Ref.\ \cite{hejhal}).
See also Ref.\ \cite{asory} for a related
discussions.

We pursue this investigation by explicit
construction, which allows us to derive, and in some sense,
generalize  all known results in a straightforward way and
provides us also the wave functions of Landau levels.
The wave functions  turn out to be the holomorphic line
bundle defined on the surface, or for the high Landau levels,
they are obtained by applying some covariant derivatives on
the holomorphic line bundle.
Actually, the holomorphic line
bundle can be defined without reference to a particular metric.
Thus although we consider  mainly
two cases for the metric,
we can make some possible interesting
generalizations to metrics of other kinds,
mainly for the case of the ground states.

It is known that the fractional quantum Hall effect (FQHE)
is related to the properties of the ground states, through
the Laughlin wave function. We will show the construction of the Laughlin
wave function on Riemann surfaces and indicate some interesting
relations with recent results appeared in the mathematical
literature\cite{fay}.

We organize the paper as follows.
In section 2, we use the Riemann-Roch theorem to derive
the degeneracy of the ground
states of a particle interacting
with a constant magnetic field
As we will see, the natural definition of a constant magnetic
field is to take it proportional to the area form.
We consider first the case of a metric which is proportional
to the Chern form of the $\theta$-function line bundle
\cite{gaume}, defined explicitly
in section $2$. Since it is expressed in terms
of the canonical holomorphic one forms, we will call
it "canonical  $\theta$-metric", abbreviated
as C$\theta$M\footnote{we understand there is not a standard
name for this metric in the literature. It is
proportional to the so called Bergman kernel,
which is also proportional  to the curvature of
the Arakelov metric.}.
In section 3, we construct the Landau levels on
the surface with the Poincar{\' e} metric.
The eigenvalues of the Landau levels and their multiplicity
are obtained.
In section 4, we continue the discussion of  section 3
to construct the wave functions of the
the Landau levels.
In section 5, we present the Laughlin wave function
on high genus surfaces with particular metrics.
The mathematical structure hidden in the Laughlin wave function
is pointed out.
In particular, we also discuss  the degeneracy of the
Laughlin states.

\section{The lowest Landau level on
Riemann surfaces with the C$\theta$M  metric}

We consider a  particle on a Riemann surface
interacting with a "monopole"   field,
that is the integral of the field strength out of the surface is different
from zero. We use the metric $ds^2=g_{z\bar z}dzd{\bar z}$
in complex coordinates and  the volume form is
$dv=[ig_{z\bar z} / 2]dz\wedge d{\bar z}
= g_{z\bar z}dx\wedge dy$.
We apply a constant magnetic field on the surface.
The natural definition
of the  constant magnetic field to the high genus Riemann
surface\cite{antoine,bolte} is
$$F=Bdv=(\partial_zA_{\bar z}-
\partial_{\bar z}A_z)dz\wedge d{\bar z},$$
with constant $B$.
Thus  $ \partial_zA_{\bar z}-
\partial_{\bar z}A_z =ig_{z\bar z}B / 2$.
The flux $\Phi$ is given by $2\pi \Phi = \int F =BV$,
where $V$ is the area of the surface and
we assume here $B>0$ ($\Phi >0$).
The Hamiltonian of an electron on the surface
under the magnetic field is given
by the Laplace-Beltrami operator,
\begin{eqnarray}
H & = & [1/ 2m \sqrt{g}]
(P_{\mu}-A_{\mu})g^{\mu \nu}\sqrt{g}(P_{\nu}-A_{\nu})
\nonumber \\
& = & [ g^{z\bar z} / m]
[(P_z-A_z)(P_{\bar z}-A_{\bar z}) \nonumber \\
& & + (P_{\bar z}-A_{\bar z})(P_z-A_z)]  \\
\label{hamil}
& = & [2g^{z\bar z}/ m]
(P_z-A_z)(P_{\bar z}-A_{\bar z})+[B/ 2m] \nonumber
\end{eqnarray}
where $g^{z\bar z}=[1 / g_{z\bar z}]$ and
$P_{z}=-i\partial_z ,\, P_{\bar z}=-i\partial_{\bar z}$
($\partial_z =(\partial_x -i\partial_y)/2$).
The inner product of two wave functions is defined as
$<\psi_1 | \psi_2 >=\int dv {\bar \psi_1 }\times \psi_2$.

$H^{\prime}=[2g^{z\bar z} / m]
(P_z-A_z)(P_{\bar z}-A_{\bar z})$
is a positive definite hermitian operator
because $<\psi | H^{\prime} |\psi > \, \geq 0$ for any $\psi$.
Thus if  $H^{\prime} \psi =0$,
then $\psi$ satisfies
$(P_{\bar z}-A_{\bar z})\psi =0$.
The solutions of this equation
are the ground states of the Hamiltonian $H$ or $H^{\prime}$,
i.e. the lowest Landau level (LLL).
The existence of the solutions of this equation
is guaranteed by the Riemann-Roch
theorem\cite{griff,griff1}.
The solutions belong to the sections of the
holomorphic line bundle under the gauge field.
The Riemann-Roch theorem tells us that
$h^0(L)-h^1(L)=deg(L)-g+1$,
where $h^0(L)$ is the dimension of the sections of
the holomorphic line bundle
or the degeneracy of the ground states of the Hamiltonian $H$,
$h^1(L)$ is the dimension of the holomorphic differential
 $(L^{-1}\times K)$ where $K$ is the canonical bundle and
$deg(L)$ is the degree of the line bundle which is equal to
the first Chern number of the gauge field, or the magnetic flux
out of the surface, $\Phi$.
When $deg(L) >2g-2$,
$h^1(L)$ is equal to zero \cite{griff}
and $h^0(L)=\Phi-g+1$.
As a consistent check,  $h^0(L)$ indeed
gives the right degeneracy
of the ground states in the case of
a particle on the sphere and torus interacting
with a magnetic-monopole field.

If the Riemann surface $\Sigma$ has $g$ ($g>0$)
handles, there exist
abelian differentials, $g$ holomorphic and $g$
anti-holomorphic closed $1$-forms, $\omega_i$ and
$\bar \omega_i$.
They are normalized by
\begin{equation}
\int_{A_i} \omega_j= \delta_{ij} \, \, ,
\int_{B_i} \omega_j=\Omega_{ij}
\end{equation}
where $A_i, B_i$ are a canonical homology basis or
closed loops around handles on $\Sigma$, and the imaginary part
of $\Omega$ is a positive matrix.
We consider the C$\theta$M metric  given by
\begin{equation}
g_{z\bar z}=
\bar \omega (Im \Omega  )^{-1}\omega,
\end{equation}
which is always greater than zero.
This metric is proportional to
the Bergman reproducing Kernel
and  it is also proportional to
the curvature of the Arakelov metric\cite{fay}.
Its most interesting feature (for our study here)
is that  it is proportional to
the Chern form of the $\theta$-function line bundle\cite{gaume},
implying the covariant derivatives match the transformation
properties of the $\theta$-function.

We note that \cite{gaume}
\begin{equation}
V=\int_{\Sigma} dxdy
\bar \omega (Im \Omega )^{-1} \omega =g.
\end{equation}
Let us take $\Phi$ equal to $\gamma g$. Because of Dirac
quantization,
$\gamma g$  must be an integer and here we assume
that  $\gamma g$ is a positive integer.
We will explicitly construct the ground states
in the case of $\gamma $ being a positive integer here
and the case of fractional $\gamma $
will be discussed in section $5.2$.
Now  we have $B=2\pi \gamma$ and
$F_{z \bar z}=\partial_zA_{\bar z}-
\partial_{\bar z}A_z
=i\pi \gamma \bar \omega (Im \Omega )^{-1}
\omega$.  We can take $A_{\bar z}=i
\pi \gamma \bar \omega (Im \Omega )^{-1}
u /2 $,  where
$u^i
= \int^z_{z_0}\omega^i $
and $A_{\bar z}=\bar A_z$
are the gauge potentials in a certain gauge.

The ground states satisfy the equation,
\begin{equation}
\bar D \Psi =
(\bar \partial +{\pi\over 2}
\gamma \bar \omega (Im \Omega )^{-1}
u )\Psi=0.
\end{equation}
Because
\begin{equation}
D(u+\Omega n+m)= K(n,m)^{-1}D(u)K(n,m)
\end{equation}
with
$$K(n,m)=\exp \gamma [S(n,m)-\bar  S(n,m)]$$
and
$$S(n,m)=(\pi /2) \bar u (Im \Omega)^{-1} (\Omega n+m),$$
thus we can choose
the boundary condition as \cite{iengole},
\begin{equation}
K(n,m)\Psi (u+\Omega n+m)=
\exp (i\alpha (n,m)) \Psi (u).
\label{boun}
\end{equation}
Generally we can take $\alpha (n,m)$
as
\begin{equation}
\alpha (n,m)=i \gamma \pi nm +2\pi i \gamma a_0m
-2\pi i  b_0n.
\end{equation}
We define the function
\begin{equation}
\Psi =\exp [-(\pi /2)
\gamma \bar u(Im \Omega )^{-1}u
+(\pi /2)
\gamma u (Im \Omega )^{-1}u ]F.
\end{equation}
Then the function $F$ satisfies the equation,
\begin{equation}
F(u+\Omega n+m)=
\exp (-i\pi \gamma n\Omega n
-2\pi \gamma nu-i \gamma \pi nm+i\alpha (n,m))
F(u).
\label{theta}
\end{equation}
The solutions of Eq.\ (\ref{theta})
are
\begin{equation}
F_i(u)=\theta \left[ \begin{array}{c}
                  a \\ b
                  \end{array} \right]
\left(\gamma u|\gamma \Omega \right)
\label{nindp}
\end{equation}
with $b=b_0$ and $a_i=(a_0)_i+l_i / \gamma$,
$l_i=0, \cdots , \gamma -1$ and $i=1, \cdots , g$.
It seems that there are $\gamma^g$ solutions.
However, from the Riemann-Roch theorem,
we know that the degeneracy is
$\gamma g -g+1$ when $\gamma >1$ (remind that
$\gamma$ is a positive integer here,  then we have
$\gamma g >2g-2$ and $h^1=0$).
We observe that
when $g=1$,  $\gamma^g = \gamma g -g+1$ but
$\gamma^g > \gamma g -g+1$
for $\gamma >1$ and $g>1$. In fact, generally
the solutions given by Eq.\ (\ref{nindp})
are not linear-independent.   Take, for example,
$F_1$  given by
\begin{equation}
F_1=\theta \left[ \begin{array}{c}
                  a_1 \\ b
                  \end{array} \right]
\left(\gamma u|\gamma \Omega \right),
\label{nindp1}
\end{equation}
then consider
$F_i / F_1$ which are the meromorphic functions on
the Riemann surface $\Sigma$.
Because $F_1$ has $\gamma g$ zeros, for example,
at $z_i, i=1, \cdots , \gamma g$,
the meromorphic functions will have possible poles at
points  $z_i$. The dimension of such
meromorphic functions is
given by the Riemann-Roch theorem, the number of the
possible poles, $\gamma g$,
minus $g-1$ in the case of $\gamma >1$,
which  is equal to $\gamma g -g+1$
and is the degeneracy of the ground states.
If $\gamma =1$, according to Eq.\ (\ref{nindp})
there is only one solution (it is possible that this solution
is identical to zero by the Riemann vanishing theorem and
thus  there will be no solutions).
We remark that if $(a_0)_i=1/ 2 \, , (b_0)_i= \gamma / 2$,
the wave functions are transformed covariantly by
the modular transformations (for the case of $g=1$,
see Ref.\ \cite{iengole} ).

\section{The Landau levels on
Riemann surfaces with the Poincar{\' e} metric}

When $g>1$, the simply connected covering
space of the surface $\Sigma$ is the upper half plane $H$
(for example, Refs.\ \cite{griff1,gaume}), and
$\Sigma$ is equal to  $H / \Gamma $, where
$\Gamma$ is the discreet subgroup of the isometry group
of $H$, isomorphic to the first homotopy of $\Sigma$.
$\Gamma$ is generated by the Fuchsian transformations around
a canonical homology basis, $\Gamma_{A_i}, \Gamma_{B_i}$
with
\begin{equation}
\prod^g_{i=1}\Gamma_{A_i}\Gamma_{B_i}
\Gamma^{-1}_{A_i} \Gamma^{-1}_{B_i}=1.
\end{equation}
The metric is given by the Poincar{\' e} metric,
$ds^2=y^{-2} (dx^2+dy^2)$, and we note that
$\int dv =\int y^{-2} dxdy =2\pi (2g-2)$ (without punctures).
In the case of the Poincar{\' e} metric $g_{z\bar z}=y^{-2}$,
the  curvature is constant:
$$
g^{z\bar z}R_{z\bar z}= -2g^{z\bar z}\partial \bar \partial ln g_{z\bar z}=
-1.
$$
We take $F=Bdv$, and thus $A_z=-i B\partial (\ln g_{z\bar z}) /2$
and the flux $\Phi$ is equal to $2B(g-1)$.
Then we define $D=\partial - (B/2) \partial \ln g_{z\bar z}$
and $\bar D= \bar \partial + (B/2) \bar \partial
\ln g_{z\bar z}$.
The Hamiltonian is (we take $m=2$ in
Eq.\ (\ref{hamil}) for the simplicity),
\begin{equation}
H = -g^{z\bar z} D \bar D +(B/ 4).
\label{hamil1}
\end{equation}
The eigenfunctions satisfy
\begin{equation}
H \Psi = E \Psi .
\end{equation}
If $\tilde z$ is another local coordinate on $\Sigma$
and the domain of $\tilde z$ intersects non-trivially the
domain of $z$, $ g_{z\bar z}dz d\bar z$ is
invariant under coordinate changes, or
\begin{equation}
g_{z\bar z}dz d\bar z =
g_{\tilde z \tilde {\bar z} }d{\tilde z}
d\tilde {\bar z}
\end{equation}
on the intersection of the domains of $z$ and $\tilde z$.
$D$ and $\bar D$ are transformed as
\begin{equation}
{\tilde D}=(d z/d
{\tilde z})U^{-1} D U \, \, ,
\tilde {\bar D}=(d\bar z/d \tilde
{\bar z}) U^{-1} \bar D U
\end{equation}
where $U(z, \tilde z)= (dz / d \tilde z)^{-B/2}
(d \bar z / d  \tilde {\bar  z}   )^{B/2}$.
The Hamiltonian is transformed as
\begin{equation}
\tilde H =U^{-1} H U,
\end{equation}
thus the wave function is transformed as
\begin{equation}
\tilde \Psi =U^{-1} \Psi
\end{equation}
or $\Psi (dz)^{B/2}(d\bar z)^{-B/2}$
is invariant under the transformation.
So we conclude that $\Psi$ is a differential form of type
$T^{\bar B/2}_{B/2}$.
Furthermore, the wave function is transformed
under the Fuchsian transformations as
\begin{equation}
\Psi (\gamma z) =u(\gamma,z) \Psi (z),
\, \, u(\gamma,z)=\nu (B, \gamma)
(cz+d)^{2B}/|cz+d|^{2B}
\label{boundaryp}
\end{equation}
where $\gamma$ is a Fuchsian group element
$\left( \begin{array}{cc}
                  a & b\\ c & d
                  \end{array} \right)$
and $\gamma z=(az+b)/(cz+d)$.
$\nu (B, -1)=e^{-i2\pi B}$ and
$u(\gamma_1 \gamma_2, z)=u(\gamma_1 , \gamma_2 z)
u(\gamma_2, z)$ are the consistency
conditions ensuring univaluedness
of the wave function on the universal
covering space \cite{hejhal}.
The boundary condition is twisted if $\nu (B, \gamma) \not= 1$.
The ground states are given by the solutions of the following
equation,
\begin{equation}
\bar D \Psi_0 =0 .
\end{equation}
The solutions of this equation are
$\Psi_0 =g^{(-B/2)}_{z\bar z} \tilde \Psi_0 $ with
$\bar \partial \tilde \Psi_0 =0$.
$\tilde \Psi_0$ belongs to a differential form of type
$T_{B}$. According to Eq.\ (\ref{boundaryp}),
$\tilde \Psi_0$ is transformed
under the Fuchsian transformations as
\begin{equation}
\tilde \Psi_0 (\gamma z) =u^{\prime}
(\gamma,z) \tilde \Psi_0 (z),
\, \, u^{\prime} (\gamma,z)=\nu (B, \gamma)
(cz+d)^{2B}
\label{boundaryp1}
\end{equation}
with $\nu (B, \gamma)$ defined in
Eq.\ (\ref{boundaryp}).
When $B=1$, $T_{B}$ is the canonical holomorphic
line bundle and $\tilde \Psi_0$ is given by
the sections of the canonical holomorphic
line bundle.
By the Riemann-Roch theorem,
we have $dim T_{B} -dim T_{1-B}=(2B-1)(g-1)$,
where  $dim T_{B}$ is the dimension of the sections
of the holomorphic bundle $T_{B}$.
The dimension (or the degeneracy of the
ground states) of $T_1$ is $g$ for
the non-twisted boundary condition and
is $g-1$ for the twisted boundary condition
($\nu \not= 1$), because  $dim T_0 =1$ for
the non-twisted boundary condition and
$dim T_0 =0$  for the twisted boundary condition.
When $B$ is an positive integer which is greater than one,
the dimension of $T_B$ is $(2B-1)(g-1)$ by
the Riemann-Roch theorem ($dim T_{1-B} =0$, as $1-B$ is negative).
When $B=1/2$, $T_{1/2}$ is $1/2$-differentials
(the spin bundle).
The dimension of $T_{1/2}$ is generically one
for the odd-spin structures and zero
for the even-spin structures.
The energy of the ground states is $B/4$. An expression for the wave
functions will be described in the next section. There we will also
indicate generalizations to the case of fractional $B$, provided that
$(2B-1)(g-1)$ is integer, and also possible generalizations to surfaces
with punctures.

Here and in the following,  we call $g=g_{z\bar z}$ for short.
We introduce the covariant derivative,
$\nabla_z$, and its Hermitian conjugate
$(\nabla_z)^{\dagger}=-\nabla^z$,
\begin{eqnarray}
& \nabla_z & : T^{l}_k \to
T^{l}_{k+1} \, \, ,  \nabla_z =g^k\partial g^{-k},  \nonumber \\
& (\nabla_z)^{\dagger} &: T^{l}_k \to T^{l}_{k-1} \, \, ,
(\nabla_z)^{\dagger} =-
g^{-l-1} \bar \partial g^{l}.
\end{eqnarray}
Note that $D$  is the covariant operator
$\nabla_z$  acting on
$T^{ \bar B /2}_{B/2}$
($\bar D = g \nabla^z$ where $\nabla^z$
acts on $T^{ \bar B /2}_{B/2}$).

Let us next discuss the higher Landau levels.
By writing
$$
 H-B/4 =-\nabla_z \nabla^z ,
$$
we notice that if
$\Psi_1$ is an eigenfunction of $H$ with eigenvalue $E_1>B/4$,
then $\Psi_1 =-\nabla_z \nabla^z \Psi_1 /\epsilon_1 $
(where $\epsilon_1 =E_1-B/4\not= 0$). Therefore
$ \Psi_1 =\nabla_z \Phi$ for some $\Phi$. Of course, since $\Psi_1$
is of the form $T^{\bar B/2}_{B/2}$, then $\Phi$ will be of the form
$T^{\bar B/2}_{B/2-1}$. Thus we have, more explicitly,
$$
\Psi_1 = (\partial - (B/2-1)\partial ln g)\Phi .
$$
Due to the property of the Poincar{\' e} metric
$\partial \bar \partial ln g=g/2$, one can easily show that
$$
-\nabla_z \nabla^z \Psi_1 ={B-1\over 2}\Psi_1 +
\nabla_z (-\nabla_z \nabla^z \Phi ).
$$
When $B\geq 1$, one can show that
$<\Psi_1|\nabla_z (-\nabla_z \nabla^z \Phi )>\geq 0$.
It is thus clear that the states of
the lowest excited level are obtained,  if
there exist $\Phi$ such that $\nabla_z \nabla^z \Phi =0$,
i.e. $\bar D\Phi= 0$. This means that $\Phi = \Phi_0 =g^{-B/2}
\tilde \Phi_0$ with $\bar \partial \tilde \Phi_0 =0$. Since $\tilde \Phi_0$
is of the form $T_{B-1}$, there exist
solutions of the equation $\bar \partial \tilde \Phi_0 =0$
for $B\geq 1$.
The energy of the
lowest excited states is thus
$$
E_1={3\over 4}B-{1\over 2}.
$$
The degeneracy of this Landau level is the  dimension of the
sections of the holomorphic bundle $T_{B-1}$
(which is equal to $(2B-3)(g-1)$ if $B>2$).
When $B<1$, there is only the  $zero'th$  "Landau level"
(the lowest Landau level).
Beyond  the  "Landau levels", little is know about the spectrum.
We will make a comment about this point at the end of the
present section.

We can generalize the above discussion to high Landau levels.
The wave function of the $k'th$ Landau level is given by
\begin{eqnarray}
\Psi_k & = & (\nabla_z)^{k}\Phi_0 \nonumber \\
&=& (\partial-(B/2-1)\partial \ln g)
(\partial-(B/2-2)\partial \ln g)  \\
&  & \cdots
(\partial-(B/2-k)\partial \ln g)\Phi_0 \nonumber
\end{eqnarray}
with $\tilde \Phi_0=g^{B/2} \Phi_0$ and
$\bar \partial \tilde \Phi_0=0$. $\tilde \Phi_0$
is a differential form of the type $T_{B-k}$. Notice that this construction
generalizes the standard construction for the harmonic oscillator.
By using the relation, which holds for the Poincar{\' e} metric,
\begin{equation}
[\nabla^z \, \, \nabla_z]T^m_n
= -(m+n)/2,
\end{equation}
one can explicitly check that $\Psi_k$ is the eigenfunction
of the Hamiltonian, with the eigenvalue as
\begin{equation}
E_k=[B(2k+1)-k(k+1)]/4.
\end{equation}
The  degeneracy of the $k'th$ Landau level
is given by the dimension of the sections of the holomorphic
bundle of   the type $T_{B-k}$, which is equal to
$(2B-2k-1)(g-1)$ when $B-k>1$.
Because the dimension of
$T_n$ is zero when $n$ is negative,
$k$ must not be greater than $B$.
Hence there is only a finite number of "Landau levels".

When $B$ is an integer, $k$ can take value
from  $0$ to $B$. When $k=B$,
the corresponding $\tilde \Psi_0$
is the (holomorphic)
differential form of the type $T_0$.
$T_0$ is a constant function on the surface. We can also include twisted
boundary conditions, which would physically correspond to the presence of
some magnetic flux through the handles.
If the boundary condition of the wave function is
the twisted one, 
there does not exist a non-zero constant function
which satisfies the twisted boundary condition.
Thus the dimension of $T_0$ is zero in this case
and there is not the $B'th$  Landau level.
When $k=B-1$, the degeneracy of this Landau level
is the dimension of the canonical bundle
$T_1$, which is equal to $g$
for the non-twisted  boundary condition
and is equal to $g-1$
for the twisted  boundary condition.
$B$ can be also an half-integer.
Then $k$ can take value from $0$ to $B-(1/2)$.
When $k=B-(1/2)$, the degeneracy of this Landau level
is the dimension of the spin bundle
$T_{1/2}$. The dimension of the
holomorphic sections of the spin bundle
generically is zero for the even-spin structures
and one for the odd ones (or for twisted ones).

In the next section,
we will show a construction of the wave functions
and we will see that it is possible to generalize the present scheme also
to
the case of $B$ fractional, provided a condition is satisfied, and to
include also "punctures" on the surface. We will
also discuss the resulting
spectrum of the Landau levels in the general case.

Of course, the "Landau levels" that we have found by the above method do not
exhaust the spectrum. In fact, when $k$ has reached the maximum value for
which $s=B-k$ is positive or zero, we can still express
$\Psi =\nabla_z^k g^{-B/2} \tilde \Phi$
where $\tilde \Phi$
is a $T_s$ differential and get an additional infinity of
levels and corresponding wave
functions by the solutions of the eigenvalue
equation for $-\nabla_z \nabla^z \tilde \Phi_n
=E_n \tilde \Phi_n$.
The corresponding
eigenvalues for $\Psi$ will be
$E={1\over 4}(B(2k+1)-k(k+1))+E_n$.
In particular, for an integer $B$,
this will relate the general solution of
our problem to the eigenvalues and eigenfunctions of the Laplacian on the
scalar (i.e. the zero forms) on the Riemann surface with the Poincar{\' e}
metric, a problem which is not completely solved and for which there exists
a vast literature ( for a recent review see Ref.\ \cite{buser}).

\section{The wave functions of the Landau levels}

In order to complete the construction of the wave functions of the
"Landau levels" of the last section,
we would like to describe $\tilde \Phi_0$, that is
the holomorphic sections of the bundle corresponding to the differential
of the type $T_s$. We will present a formula for the determinant
$\det h_i(z_j)$,
where $h_i$ are the independent holomorphic sections and
$i,j$ run over the degeneracy of the Landau level. To get a particular
wave function,  it is of course enough to consider this determinant as a
function of a particular $z$, fixing arbitrarily the remaining ones.
We have anticipated from the Riemann-Roch theorem that the degeneracy
is $N=(2s-1)(g-1)$, for $s$ integer or half integer greater than $1$,
the cases $s=1$, $s=1/2$, $s=0$ corresponding respectively to the
$g$ abelian differential, to the holomorphic spin structure(s) and
to the constant respectively, as recalled above.
The following formula does not make
reference to any particular metric, as the notion of holomorphic
differentials is introduced in a metric independent fashion.

The formula can be read from Ref.\ \cite{iengo}, and it has been
obtained in a contest of String theory, following the work of
Knizhnik\cite{kn}. It is:
\pagebreak
\begin{eqnarray}
\det h_i(z_j) & = & \theta \left[ \begin{array}{c}
                  a \\ b
                  \end{array} \right]
\left(\sum u_i -(2s-1)\sum_{i=1}^{g-1}r_i \right)
\prod^{g-1}_{i=1}
\left( \nu_0(z_i) \right)^{2s-1} \nonumber \\
&  & \times \prod^N_{i<j}E(z_i, z_j)/\prod^N_{i=1}\prod^{g-1}_{l=1}
\left( E(z_i, r_l) \right)^{2s-1},
\label{det}
\end{eqnarray}
where $\nu_0(z)$ is a holomorphic half-differential with $g-1$ zeros,
corresponding to an arbitrary (but fixed) odd characteristic
and $\nu_0 (r_i)=0, \, \, i=1, \cdots , g-1$. The symbol $E(z_i,z_j)$
denotes the "prime form",
which is a $-1/2$ differential and it
has only one zero for $z_i=z_j$ (for a review of definitions and
transformation properties of the theta functions, prime forms etc.,
see for  instance Ref.\ \cite{ed}).
Recall that $h_i(z)$ is the solution of the following equation,
\begin{equation}
\bar D g^{-s/2} h(z)=g^{-s/2}\bar \partial h(z)=0
\label{anym}
\end{equation}
with  $\bar D =\bar \partial + (s/2) \bar \partial
\ln g$.
What we discuss now is valid for any metric, not only
for the Poincar{\' e}  metric.
The above equation implies that
\begin{equation}
\bar D_{z_k} \prod^N_{i=1}g^{-s/2}(z_i,\bar z_i)
\det h_i(z_j)=0.
\label{anymd}
\end{equation}
Thus with respect to  the coordinate $z_i$,
the function $\det h_i(z_j)$ shall be a form of type $T_s$,
which we can show directly from Eq.\ (\ref{det}).
Because the prime form is a $(-1/2)$ form
and $\nu_0$ is a $1/2$ form,
\begin{equation}
{1\over 2}(2s-1)-{1\over 2}\left( (2s-1)(g-1)-1 \right)
+(2s-1)(g-1)=s,
\end{equation}
which is the type of the form with respect to  every coordinate $z_i$.

It is seen that it is indeed holomorphic,  the zeroes in the denominator
are canceled by corresponding zeroes in the numerator.

The variable in  the theta function is
$(\sum u_i -(2s-1)\sum_{i=1}^{g-1}r_i)$.
We write the theta function in this way because
the phase obtained by moving the coordinates around
handles will be independent
on the zeros $r_i$ in this format
(this is only for the convenience).
We can also absorb it into the characteristics.
In contrast to the case when the magnetic field is proportional
to the C$\theta$M metric, in the present case
the gauge potential $A_z=-i s\partial (\ln g) /2$ is
single-valued around the handles. Thus the boundary condition
shall be
\begin{equation}
h_i(u+\Omega n+m)=\exp (\alpha n+\beta m)h_i(u).
\end{equation}
The above equation implies that
\begin{equation}
\det h_i(u_j +\Omega n+m)=\exp (\alpha n+\beta m)
\det h_i(u_j).
\label{bound1}
\end{equation}
The right side of Eq.\ (\ref{det}) indeed
transforms around the handles in the way
described by Eq.\ (\ref{bound1}). The values of $\alpha,\beta$ are
determined by
the type of the odd characteristic $\nu_0$ and the values of the
characteristics $a,b$ of the theta function. Therefore
by using Eq.\ (\ref{bound1}), the characteristics
$a$ and $b$ in Eq.\ (\ref{det})
can be fixed uniquely.

By Eq.\ (\ref{boundaryp1}),
it is easy to show that $\det h_i(z_j)$ shall transform in the
case of the Poincar{\' e} metric
under the Fuchsian transformations as
\begin{equation}
\det h_i(\gamma z_j) =u^{\prime}(\gamma,z) \det h_i(z_j),
\label{boundp}
\end{equation}
where $u^{\prime} (\gamma,z)$
is given by Eq.\ (\ref{boundaryp1}).
We can check that Eq.\ (\ref{det}) transforms
in the way of Eq.\ (\ref{boundp})
in the case of the Poincar{\' e} metric.
The transformations of the prime form and half -differential
under the Fuchsian transformations  can be found in Ref.\ \cite{fay}.
$\nu (B, \gamma)$ corresponds to the boundary condition parameters
and the characteristics $a$ and $b$ in Eq.\ (\ref{det}) are fixed by
$\nu (B, \gamma)$.

The formula Eq.\ (\ref{det}) makes sense also for $s$ fractional, provided
that $N=(2s-1)(g-1)$ is a positive integer, giving the multiplicity of the
level (apart from the particular case $s=1$ recalled above), for
generic characteristics $a,b$ corresponding to twisted boundary conditions.
The value of $s$ for this case is $s>1/2$. The dual line bundle will be
a form of the type $T_v$ with $v=1-s<1/2$. If $s>1$, since there
are no holomorphic $T_v$ forms with $v<0$, $N$ is the multiplicity for any
characteristic. In the case where $1>s>1/2$,
then $1/2>v>0$, and for the generic
moduli of the Riemann surface,
there will be  one holomorphic $T_v$ form
for some characteristics.
Thus, for these characteristics, the multiplicity
of the level corresponding to $T_s$
will be $N+1$. The formula for the wave function would be in
this case a generalization of
Eq.\ (\ref{det})\cite{fay}. For those characteristics
there exists also a Landau level, generically with multiplicity $1$,
for those $s<1/2$ for which $(1-2s)(g-1)$ is a positive integer.

Since $s=B-k$ (recalled from the last section),
this means that we find
Landau levels provided that $(2B-1)(g-1)$
is a positive integer,
or, with some characteristics,
that $(1-2B)(g-1)$ is a nonnegative integer (remember that we
assume $B>0$). Therefore the general condition
for the existing of Landau levels
is that $2B(g-1)$ is
integer, that is the Dirac quantization condition.
We remind that the energy of the level is
$$
E_k={1\over 4}(B(2k+1)-k(k+1))
$$
corresponding to $s=B-k$. Thus we see that
the Landau level of maximal energy is obtained
for the value of $k$ which is nearest to $B-1/2$. If $B$ is integer then
the maximal energy is $B^2/4$, if $B$ is half integer the maximal energy is
${1\over 4}(B^2+1/4)$. If $B$ is another allowed
fraction, the maximal energy
is intermediate between the previous two.

This discussion about Eq.\ (\ref{det})
can be further generalized to the case of "punctures",
which formally corresponds
to the possibility of allowing poles for
$\Phi_0$ at some points of the surface, with the understanding that those
points are infinitely far (with the Poincar{\' e} metric) from any other
points. This means that the puncture
can be taken at infinity or on the real
axis in the upper half-plane, the surface making a narrow cusp there such
that the area is still finite.  Thus this discussion makes
now use of a particular metric on the Riemann surface.
Quantum mechanically we require the wave function to be normalizable
and (taking the puncture at infinity) this implies for a differential
$T_s$ requiring $(y^2)^{s-2}|T_s|^2$ to be integrable in $y$
for $y \to \infty$.  This means
that the poles of $T_s$ can be of order $s$ at most,
since a pole of order $r$ at $z=\infty$ would imply
$\lim_{y\to \infty} T_s \sim y^{r-2s}$ as it is seen by the appropriate
change of chart.
If we allow for punctures at say $w_1,...,w_n$, we have the freedom of
generalizing
Eq.\ (\ref{det}) by multiplying the r.h.s by
$\prod_i \prod_l (E(z_i,w_l))^{-s}$ and subtracting $(s\sum w_l)$ from
the argument of the theta function (this insures that (\ref{bound1})
continues to hold). This construction gives again a $T_s$ differential
provided now $N=(2s-1)(g-1)+ns$, which is the new multiplicity of the level
(we should be aware that  $N=(2s-1)(g-1)+ns$ may not  be
true in the case of $s\leq 1$. See the previous discussion).

Finally, a further generalization
could consist in allowing for some twists
on the punctures, corresponding to considering
branch points rather than poles at $w_l$.

\section{The Laughlin wave function
on Riemann surfaces}

\subsection{The Laughlin wave functions
in the constant field on the surface with the
Poincar{\' e} metric}

In the present section, we will  discuss the Laughlin wave functions
in the constant field on the surface with the
Poincar{\' e} metric.
In the next subsection, the Laughlin states
in the constant magnetic field  will be worked out  in the case
of the magnetic field which is proportional to the C$\theta$M
metric. The mathematical structure behind
the Laughlin wave functions  will be discussed in the end of this
subsection.  We shall remark that the following discussions
can be generalized to the case of the magnetic field being
proportional to the curvature, if we take   the Hamiltonian
in a special ordering,
\begin{equation}
H = -g^{z\bar z} D \bar D
\end{equation}
with $A_z=-i B\partial (\ln g) /2$ and $g$ is an arbitrary metric.
The magnetic field is a constant one
if the  curvature is constant,  for example, in the case of
the Poincar{\' e} metric (see the last section).
The generalization is straight forward and we will not discuss
it here.

Following section $3$,
we take $F_{z\bar z}=iB\bar \partial \partial \ln g$,
$A_z=-i B\partial (\ln g) /2$ and $g>1$.
The ground states satisfy
the equation, $\bar D \Psi_0=0$ and the solution of the equation
is $\Psi_0 =g^{-B/2} \tilde \Psi_0 $ with
$\partial \tilde \Psi_0 =0$. $\Psi_0$ is
$T_B$ differentials.
$h_i(z)$ are the solutions of the equation
$\partial \tilde \Psi_0 =0$.
In the FQHE,
the magnetic field applied is very strong.
Thus $B$ is a very large number and
the number of the sections of the
holomorphic $T_B$ differentials
is equal to $(2B-1)(g-1)$.
If  the ground states are completely filled,
which corresponds to the case of the quantum Hall state
with filling $\nu =1$, the wave function of the
quantum Hall state is given by
$\Psi_{JL} =\prod_{i=1}
g^{-B/2}(z_i\bar z_i)
\det h_i(z_j)$ where $i=1, \cdots , (2B-1)(g-1)$
($\Psi_{JL}$ stands for the Jastrow-Laughlin type wave function).

A formula for $\det h_i(z_j)$ has
been shown and discussed in the previous
section, see Eq.\ (\ref{det}).
For any quantum Hall state, we write
$\Psi_{JL} =\prod_{i=1} g^{(-B/2)}(z_i\bar z_i)\Psi^{\prime}_J$
and $\Psi^{\prime}_J$ is a holomorphic function of
the coordinates of any electrons, as $\Psi_{JL}$ satisfies
the equation $\bar D_{z_i}\Psi_{JL}=0$
(we have this equation because every electron stays in the LLL).
The boundary condition,
Eq.\ (\ref{boundaryp}) or Eq.\ (\ref{boundaryp1})
for the single particle implies a boundary condition
on the many-body wave function (here is the wave function of the Hall
state),
\begin{equation}
\Psi^{\prime}_{JL} (\gamma z_j)
=u^{\prime} (\gamma,z) \Psi^{\prime}_{JL} (z_j),
\label{boundpp}
\end{equation}
with $u^{\prime} (\gamma,z)$ given by
by Eq.\ (\ref{boundaryp1}).
Furthermore, by using Eq.\ (\ref{boundpp}), the characteristics
$a$ and $b$ in Eq.\ (\ref{det})
can be fixed uniquely. Hence the degeneracy
of the Hall state at filling $\nu =1$ is one
(from the physical points of view, there is only one way
to completely fill the lowest Landau levels).

If  the filling  is equal to $1/m$,
we make an Ansatz for the wave function,
$\Psi_{JL} =\prod_{i=1}
g^{(-B/2)}(z_i\bar z_i)\Psi^{\prime}_{JL}$,
\begin{eqnarray}
\Psi^{\prime}_{JL} & = & \theta \left[ \begin{array}{c}
                  a \\ b
                  \end{array} \right]
\left(m\sum u_i -Q\sum_{i=1}^{g-1}r_i |
m\Omega \right)
\prod^{g-1}_{i=1}
\left( \nu_0(z_i) \right)^{Q_1} \nonumber \\
&  & \times \prod^N_{i<j}
\left( E(z_i, z_j) \right)^m /\prod^N_{i=1}
\prod^{g-1}_{l=1}
\left( E(z_i, r_l) \right)^{Q_2},
\label{det1}
\end{eqnarray}
where $N$ is the number of the electrons.
$Q_1$ must be equal to $Q_2$, otherwise, there will be
singularities at $r_i$. The way we write
the theta function  in Eq.\ (\ref{det1})
is based on the intuition from the
wave function on the torus.
Moreover, under the Fuchsian transformation,
Eq.\ (\ref{det1}) shall be transformed
as Eq.\ (\ref{boundpp}).
This   implies that
$Q_1=Q_2=2B-m$ and $m(N-1+g)=2B(g-1)=\Phi$.
And we take $Q=Q_1$ for the same
reason in the case of $\nu =1$.
The characteristics $a$ and $b$ are determined by the boundary
condition, which gives
\begin{equation}
b=b_0, \, \, a=a_0+l/m,  \, \, l_i=1, \cdots , m, \, \,
i=1, \cdots , g,
\end{equation}
where $a_0$ and $b_0$ depend on the boundary value parameters
$\alpha$ and $\beta$.
The above  equation does not imply that
the degeneracy of the Laughlin wave functions is $m^g$.
According to the second section,
the linear-independent number of the functions
\begin{equation}
\theta \left[ \begin{array}{c}
                  a \\ b
                  \end{array} \right]
\left(m\sum u_i -Q\sum_{i=1}^{g-1}r_i |
m\Omega \right)
\label{thetaf}
\end{equation}
is  $mg-g+1$
and   thus the degeneracy is actually $mg-g+1$

There is a deep mathematical structure behind the Laughlin wave
functions.  In the case of $\nu =1$,
we know the wave function is given
by the determinant of the sections of
the holomorphic line bundle.
In the case of $\nu =m $ with $m$ greater than $1$,
the Laughlin wave functions is given by
the determinant of the sections of
the holomorphic rank-$m$  vector bundle.
The discussion about the determinant
of the sections of the holomorphic
vector bundle can be found in Ref.\ \cite{fay}.

We consider a line bundle with connection
$A^{\prime}_z=-iB^{\prime}\partial \ln g /2$
(thus  $\Phi^{\prime} =2B^{\prime}(g-1)$
is the corresponding magnetic flux and is a positive integer).
We take $B^{\prime}=B/m$ for the reasons to be seen later.
Under the Fuchsian transformations,
the transformations of the line bundle
are given by Eq.\ (\ref{boundaryp}) with the replacement of
$B$ by $B^{\prime}$.
We also introduce a flat vector bundle of rank-$m$
with a flat connection. The flat vector bundle
$\Psi (\gamma z)_k$ transforms
under the Fuchsian transformations as
$$\Psi (\gamma z)_k=\sum^{m}_{i=1} \chi(\gamma )_{kj} \Psi (z)_j,$$
where $\chi(\gamma )$ is a constant matrix.
We define a vector bundle $E$ as the tensor product of the
line bundle and the flat vector bundle.
The vector bundle then is transformed
under the Fuchsian transformations as,
\begin{equation}
\Psi (\gamma z)_k =\sum_{l=1}^{m} u(\gamma,z)_{kl} \Psi (z)_l,
\, \, u(\gamma,z)=\nu (B^{\prime}, \gamma)
(cz+d)^{2B^{\prime}}/|cz+d|^{2B^{\prime}}
\label{boundvec}
\end{equation}
where $\gamma$ is a Fuchsian group element
and $\nu (B^{\prime}, \gamma)$ is now a $m\times m$ matrix.
Following the discussion about the line bundle,
we shall have
$ \nu (B^{\prime}, -1)=e^{-i2\pi B^{\prime}}$ and
$u(\gamma_1 \gamma_2, z)=u(\gamma_1 , \gamma_2 z)
u(\gamma_2, z)$ are the consistency
conditions ensuring univaluedness
of the wave function on the universal covering space.
The holomorphic sections of the vector bundle
is the solution of the following equation,
\begin{equation}
(P_{\bar z_i}-A^{\prime}_{\bar z} ) {\vec \Psi} (z) =0.
\label{madet11}
\end{equation}
where ${\vec \Psi} (z)$ is a $m$-dimension  vector
with the component $\Psi (z)_k$.
The degree of the vector bundle is $deg(E)=m \times \Phi^{\prime} =\Phi $.
We assume here $\Phi^{\prime}$
or $ \Phi $ is much larger than one
(because  we apply a very strong magnetic field in the FQHE),
$h^1$ is zero by the Kodaira vanishing theorem,
which states that
there do not exist the sections of the
one-form holomorphic vector bundle
or the  holomorphic vector bundle $E^{-1}\times K$
where $E$ is the vector  bundle and $K$ is the canonical
bundle.
It is possible to see that the vector bundle $E^{-1}\times K$
is negative\cite{coates}.
Then by the Kodaira vanishing theorem,
the dimension of the  sections of the
holomorphic vector bundle $E^{-1}\times K$
is zero\cite{coates}. From the Riemann-Roch theorem
for the vector bundles, the dimension of  the sections of the
holomorphic vector
bundle is $h^0(E)= h^1(E)+deg(E)+m(1-g)=\Phi +m(1-g)$.
Suppose  the basis of the vector bundle
is given by ${\vec \Psi}_i(z)$
with $i=1, \cdots, h^0(E)$,  we can construct a determinant,
$\det {\vec \Psi}_i(z_j)$ with  $j=1, \cdots , N$
and $N=h^0(E)/m$.

One can show that
\begin{equation}
\bar \partial \det (\Lambda_{i,j})=
\sum_k \det (\Lambda^{\prime}_{i,j}(k))
\label{miden}
\end{equation}
where the matrix $\Lambda^{\prime}_{i,j}(k)$
is given by $\Lambda^{\prime}_{i,j}(k)
=\Lambda_{i,j}, \, \, i\not= k $
and $\Lambda^{\prime}_{k,j}(k)= \bar
\partial \Lambda_{k,j}$.
By using Eq.\ (\ref{miden}) and Eq.\ (\ref{madet11}),
we can prove that
\begin{equation}
(P_{\bar z_i}-mA^{\prime}_{\bar z}) \det {\vec \Psi}_i(z_j)=0.
\label{madet}
\end{equation}
$\det {\vec \Psi}_i(z_j)$ is an anti-symmetric functions
with respect to the interchange of
any coordinates $z_i$ and $z_j$.
Thus the  above equation shows  that $\det {\vec \Psi}_i(z_j)$
is the  wave function of the electrons
interacting  with the magnetic field
$mA^{\prime}_z=A_z$.
The flux out of surface of this magnetic field $A_z$
is equal to $\Phi =m \Phi^{\prime}$.
We shall show that $\det {\vec \Psi}_i(z_j)$ is a
Laughlin type wave function.
Furthermore, we have a relation $m(N-1+g)=\Phi$
and the above discussions offer a mathematical
explanation of this relation for the Laughlin states.
This relation had been used to calculate the spin of the quasiparticle
in Ref.\ \cite{li} and the value of the spin is
found to be topological independent.

Now we shall prove that $\det {\vec \Psi}_i(z_j)$
is a Laughlin type wave function.
By Eq.\ (\ref{boundvec}),  one can also show that
\begin{equation}
\det {\vec \Psi}_i(\gamma z_j)=\det (\nu (B, \gamma))
(cz_j+d)^{2B}/|cz_j+d|^{2B} \det {\vec \Psi}_i(z_j)
\label{vectran}
\end{equation}
with $\det (\nu (B, -1))=e^{-i2\pi B}$.
Thus $\det {\vec \Psi}_i(z_j)$
transforms in the same way as the Laughlin wave function
$\Psi_{JL}$. Moreover,
when $z_i \to z_j$, we can easily show
that $\det {\vec \Psi}_i(z_j)  \to (z_i-z_j)^m$.
Thus the function obtained by taking the ratio
of $\det {\vec \Psi}_i(z_j)$ with
$$
\prod^{g-1}_{i=1}
\left( \nu_0(z_i) \right)^{2B-m}
\prod^N_{i<j}
\left( E(z_i, z_j) \right)^m /
\prod^N_{i=1}g^{B/2}(z_i,\bar z_i)
\prod^N_{i=1} \prod^{g-1}_{l=1}
\left( E(z_i, r_l) \right)^{2B-m},
$$
has no poles. By using Eq.\ (\ref{madet11}), one can show
that this function  is an holomorphic
function of coordinates $z_i$.
Furthermore,  by using Eq.\ (\ref{vectran}),
we find that this function
transforms exactly in the way as the theta functions
in  Eq.\ (\ref{thetaf}).
This  holomorphic function
must be  equal to one of  the theta functions
in  Eq.\ (\ref{thetaf}). Thus we complete
our prove that $\det {\vec \Psi}_i(z_j)$ is a
Laughlin type wave function.
Similar  arguments, for example in
Ref.\ \cite{verlinde},  had been used in proving
some identities.

{}From the discussion of the previous section,
$B$ can be fractional in the above formulae
(the above Laughlin wave functions make sense
even $B$ is fractional),  but $N$ and $\Phi$ must be integers.

We comment that the construction of the Laughlin wave function
by the determinant of the sections of the holomorphic vector bundle
can also apply to the case of $g=0,1$.
It seems to us that the degeneracy of  Laughlin wave functions
is related to the different choice of
the basis of the sections of the
holomorphic vector bundle.
Finally we shall point out that the hierarchical wave function
on the Riemann surface can also be constructed
by following the method developed in Ref.\ \cite{lidpt}

\subsection{The Laughlin wave functions
in the constant field on the surface
with the C$\theta$M metric}

Following the first section, we take
$F_{z \bar z}=\partial_zA_{\bar z}-
\partial_{\bar z}A_z
=i\pi \gamma \bar \omega (Im \Omega )^{-1}
\omega$.   $A_{\bar z}=i
\pi \gamma \bar \omega (Im \Omega )^{-1}U /2 $.
The ground states are given by
\begin{eqnarray}
\Psi_i & = & X(u)F_i(u) , \nonumber \\
X(u)& = & \exp [-(\pi /2)
\gamma \bar u(Im \Omega )^{-1}u
+(\pi /2)  \gamma u (Im \Omega )^{-1}u ],
\end{eqnarray}
where $i=1, \cdots , \gamma g-g+1$ and we assume here that
$\gamma >1$ and $\gamma$ is an integer.
We remark that the following formula for the wave function
is true also in the case of fractional $\gamma$
and $\gamma g>2g-2$ (remind that $\gamma g$
is always an integer).
$F_i$ are the linear independent solutions
of Eq.\ (\ref{theta}).
The wave function of the electrons when the first Landau
levels (or ground states) are completely filled
(the filling is equal to $1$ in this case) is then
\begin{equation}
\Psi_{\nu =1} =\det \left( X(u_j)F_i(u_j) \right)
=\prod_i X(u_i)\det \left( F_i(u_j) \right).
\end{equation}
$\det \left( F_i(u_j) \right)$ can be
calculated even we do not know
how to select $F_i(u)$, the linear independent solutions
of Eq.\ (\ref{theta}).
According to Ref.\ \cite{fay},
$\det \left( F_i(u_j) \right)$ is equal to
\begin{eqnarray}
\theta \left[ \begin{array}{c}
                  a \\ b
                  \end{array} \right]
\left(\sum u_i \right)
\prod^N_{i<j}E(z_i, z_j) f(z_1, \cdots, z_N),
\end{eqnarray}
where $N=\gamma g-g+1$, $f(z_i,z_j)=f(z_j,z_i)$
and $f(z_i)$ has no zeros with respect to
any coordinates.
The function $f(z_i)$ can be determined by the boundary
condition. From Eq.\ (\ref{theta}), we have
\begin{eqnarray}
\det \left( F_i(u_j+\Omega n+m) \right)
& = &
\exp (-i\pi \gamma n\Omega n
-2\pi \gamma nu_j \nonumber \\
& & -i \gamma \pi nm+i\alpha (n,m))
\det \left( F_i(u_j) \right).
\label{thetas}
\end{eqnarray}
Remarkably, we find that  $\det F_i(u_j)$ is given by
the same formula as that in Eq.\ (\ref{det}),
that is $\det F_i(u_j)=\det h_i(z_j)$.  One has only to
replace  $2s$ by $\gamma$ and take $N=\gamma g-g+1$
(instead of $N=2s(g-1)-g+1$, as it was
in Eq.\ (\ref{det})).
One can then verify that Eq.\ (\ref{thetas}) is
indeed satisfied.
Moreover,  the characteristics $a$ and $b$ are uniquely fixed
by the boundary condition.
We are not surprised that $\det F_i(u_j)$ is given by
Eq.\ (\ref{det}) because both of them are the determinants
of the holomorphic sections of  bundles.

If filling $\nu =1/m$,
the Laughlin wave functions
are given by, $\Psi_{JL}=\prod_i X(u_i)\Psi^{\prime}_J$
with $\Psi^{\prime}_J$ taking the same form as
Eq.\ (\ref{det1}).
However,  now we shall take
$Q=Q_1=Q_2=\gamma -m$, $\gamma =\Phi/g$
compared with $Q=2B-m$ and $B=\Phi /2(g-1)$
in the case of the Poincar{\' e} metric.
Always, we  have the relation $m(N-1+g)=\Phi$.
The wave functions shall satisfy the boundary condition
\begin{eqnarray}
\Psi^{\prime}(u_i+\Omega n+m)
& = &
\exp (-i\pi \gamma n\Omega n
-2\pi \gamma nu_i \nonumber \\
& & -i \gamma \pi nm+i\alpha (n,m))
\Psi^{\prime}(u_i) .
\label{thetasnn}
\end{eqnarray}
and the characteristics $a$ and $b$ in Eq.\ (\ref{det1})
can be fixed by the boundary condition, e.g.,
Eq.\ (\ref{thetasnn}).
The characteristics are given
$b=b_0, \, \, a=a_0+l/m. \, \, l_i=1, \cdots , m$
and $i=1, \cdots , g$,
where $a_0$ and $b_0$ depend on the boundary value parameters
$\alpha (n, m)$. It is easy to see that the degeneracy
of the Laughlin wave functions is also equal to $mg-g+1$.

By making continuation of $\gamma$,
$\gamma$ can be fractional, although
$N$ and $\Phi$ are always integers.
Now we can show how to obtain the ground state wave functions of
a particle interacting with a constant magnetic field
with the C$\theta$M metric in the case of fractional $\gamma$.
When $\gamma g >2g-2$ and $\gamma$ being fractional,
the degeneracy of the ground states is still given by
$N=\gamma g -g+1$ and
the wave function of the electrons
at filling $\nu =1$  is still given by
Eq.\ (\ref{det}) multiplied by $\prod_i X(u_i)$.
The wave function of a single particle
can be  obtained by fixing the coordinates of other particles
in  the wave function of the electrons at filling $\nu =1$
in this case.

\subsection{The degeneracy of the Laughlin state,
a general discussion}

We have show that the degeneracy of some Laughlin states
in the last two subsections is $mg-g+1$ and we will try
to show here that, generally,  the degeneracy of
Laughlin states
is $mg-g+1$ under some
reasonable assumptions.
The Laughlin type wave function is the
many  particle wave function which looks like
$\Psi_{JL}=F(z_1, \cdots,  z_N)\prod_{i<j}[f(z_i,z_j)]^m$,
where $f(z_i,z_j)=-f(z_j,z_i)$ and it is a
function on holomorphic coordinates $z_i$
and $z_j$. Furthermore when $z_i$ approaches $z_j$,
$f(z_i,z_j)$ is proportional to $z_i-z_j$ and only when
$z_i=z_j$, $f(z_i,z_j)=0$.
The only function with those properties on Riemann surfaces  is
the prime form function $E(z_i, z_j)$.
Because every particle  stays in  the lowest Landau level, so
$(P_{\bar z_i}-A_{\bar z_i})\Psi_{JL} =0$,
Hence  the Laughlin wave function  will be
\begin{equation}
\Psi_{JL}=\prod_{i=1}^{N} \exp [\int A_{\bar z_i} d \bar z_i]
\prod^N_{i<j}[E(z_i,z_j)]^m
\times F^{\prime}(z_1, \cdots,  z_N)
\end{equation}
where $F^{\prime}(z_i)$ is a function of holomorphic
coordinates $z_i$.
$F^{\prime}$ shall be determined by the boundary condition
on the surface which is compatible with the Hamiltonian.
Suppose that the magnetic field is smooth enough,
we expect that
$\Psi_{JL}$ has no poles and
$\exp [\int A_{\bar z_i} d \bar z_i]$ has no zeros and poles.
We have shown that $m(N-1+g)=\Phi$ for the Laughlin states
in the last two subsection and offered a mathematical
explanation of this relation.
Thus we can assume that
$m(N-1+g)=\Phi$ is true for any Laughlin states.
Because the degree is equal to $\Phi$, so
$\Psi_{JL}$ has $\Phi$ zeros with respect to every coordinate.
The number of zeros (counting the order of the zeros) in the
function $\prod_{i<j}[E(z_i,z_j)]^m$
is $m(N-1)$ with respect to $z_i$.
So the remaining $\Phi - m(N-1)=mg$
zeros are contained in the function $F^{\prime}$.
Suppose we have solutions $F_i^{\prime}$,
thus $G_i={F_i^{\prime} \over F_1^{\prime}}$
is a meromorphic function with respect to  $z_i$, where
$F_1^{\prime}$ is one of the solutions
(possibly some zeros in $F_i^{\prime}$
cancel some zeros in $F_1^{\prime}$).
{}From the previous discussion,
we make an assumption that $G_i$ is a function of the center
coordinate $\sum_i z_i$.
Because the meromorphic function can be always given by one
$\theta$ function divided by another
$\theta$ function, so the meromorphic function
with  poles at the points which lie on some  zeros
of the function $F_1^{\prime}$ is given by
\begin{equation}
G_i={\frac {\theta {a_i\brack b_i}(mu|m\Omega)}
{\theta {a_1\brack b_1}(mu|m\Omega)}}
\end{equation}
where now $u=\sum_i \int^{z_i} \omega$ and
${\theta {a_1\brack b_1}(mu|m\Omega)}$
has same zeros as $F_1^{\prime}$.
By Riemann-Roch theorem, the number of  such
linear independent meromorphic  functions
is $mg-g+1$ for $m > 1$ ($m=1$ is the case of the integer
QHE  and the degeneracy of the Hall state is one).
Thus the number of  the linear
independent functions $F_i^{\prime}$
or the degeneracy of the Laughlin states is
$mg-g+1$. However in Refs.\ \cite{gread,niu},
it was pointed out that the degeneracy of the Laughlin
states is s $m^g$  on the  surface with $g$ handles.
In Ref.\ \cite{niu}, Wen and Niu analyzed the Chern-Simons effective theory
of the FQHE to get the degeneracy of the Laughlin states.
In the Chern-Simons theory, there are so called large components
of gauge fields (for example, Ref.\ \cite{bos})
and one part of the wave function
is $F^{\prime}_l({\cal A})=
{\theta {a_0+{l\over m} \brack b_0}(mu+m{\cal A}|m\Omega)}$,
where ${\cal A}$ is the large component of the gauge field,
$a_0+{l\over m}, b_0$ take values in ${\bf R}^g / {\bf Z}^g$
and $a_0, b_0$ is dependent on the boundary condition.
The phase space of ${\cal A}$ is Jacobian
variety ${\bf C}^g / {\bf Z}^g+\Omega {\bf Z}^g$.
Because ${\cal A}$ is now a dynamical variable,
all $m^g$ functions $F^{\prime}_l({\cal A})$ are independent
with each other. Thus the degeneracy of the Laughlin states is
$m^g$. However if we suppose that ${\cal A}$ is a constant
vector, the number of the linear independent functions
among $F^{\prime}_l({\cal A})$ is $mg-g+1$.


\section{Acknowledgements}

We would like to thank Professors B. Dubrovin and
K.S. Narain for many useful conversations.
The work is partially supported by EEC, Science
Project SC1$^*$-CT92-0789.
\pagebreak

\font\pet=cmr10 at 10truept
\font\bf=cmbx10 at 10truept

\pet

\end{document}